\documentclass[twocolumn,letter]{jpsj3}

\usepackage{txfonts}
\usepackage{graphicx}
\usepackage{dcolumn}
\usepackage{bm}
\usepackage[usenames,dvipsnames]{xcolor}
\usepackage{ulem}
\usepackage{amsmath}
\usepackage{braket}
\usepackage{enumerate}
\usepackage{arydshln}
\usepackage{multirow}

\title{Chirality/Axiality-Induced Axiality/Chirality via Surface Polarization}
\author{Satoru Hayami, Rikuto Oiwa, and Akane Inda}
\inst{
   Graduate School of Science, Hokkaido University, Sapporo 060-0810, Japan
 }
\abst{ 
In condensed matter physics, a broad spectrum of physical characteristics, such as chirality, axiality, and polarity, arises as a direct consequence of the underlying symmetry of the system. 
We here theoretically investigate the effective coupling between chirality and axiality at their domain boundaries, mediated by polarity.
Based on symmetry considerations and model analyses, we propose the concept of chirality-induced {axiality via surface polarization}, which refers to a phenomenon where the handedness of chirality selects an axial moment with a particular orientation by lowering its energy at the surface. 
We further establish the inverse process, termed axiality-induced chirality via surface polarization, whereby axiality in turn dictates the preferred chirality. 
These reciprocal couplings open a new pathway for stabilizing single-domain states of chirality and axiality. 
They further imply interfacial functionalities, including the selective adsorption of chiral and axial molecules.
}

\begin{document}
\maketitle

Symmetry, and its breaking, lies at the heart of condensed matter physics, giving rise to a wide range of electronic, magnetic, and structural properties. 
Chirality, the property of an object being distinguishable from its mirror image, is characterized by the breaking of both spatial inversion and mirror symmetries while preserving time-reversal symmetry~\cite{kelvin1894molecular, kelvin2010baltimore, Barron1986jacs_asymmetric_systhesis, Barron_2004, bousquet2025structural}. 
This unique form of symmetry lowering makes chirality a fertile source of diverse quantum phases and physical phenomena, including the Edelstein effect~\cite{yoda2015current, furukawa2017observation, Furukawa_PhysRevResearch.3.023111}, the electrical magnetochiral effect~\cite{Rikken_PhysRevLett.87.236602, yokouchi2017electrical, Aoki_PhysRevLett.122.057206, rikken2022dielectric}, unconventional superconductivity~\cite{Togano_PhysRevLett.93.247004, mitsuhashi2010superconductivity, qin2017superconductivity, carnicom2018tarh2b2, nakajima2023giant, nakajima2024superconductors}, magnetic skyrmions~\cite{Muhlbauer_2009skyrmion, yu2010real, nagaosa2013topological, Tokura_doi:10.1021/acs.chemrev.0c00297, hayami2024stabilization}, and chirality-induced spin selectivity (CISS)~\cite{B.Gohler_nat_2011_CISS, O.Ben_nat_2017_CISS, K.Michaeli_PNAS_2019_CISS, Suda_natcom_2019_CISS, A.Inui2020prl_CrNb3S6, R.Neeman_ACR_2020_CISS, shitade2020geometric, Waldeck2021aplmat_CISS, Bloom_2024_chemrev_CISS}. 
Beyond its fundamental significance, chirality also plays a pivotal role in practical applications. 
In particular, the ability to control and manipulate handedness is central to enantioselective chemistry, molecular recognition, and pharmaceutical design, where selective interactions between left- and right-handed states govern functionality and efficiency~\cite{dalko2001enantioselective, barron2008chirality, nafie2011vibrational, Koyel_2018_science_enantioseparation, Tassinari_ChemSci_2019_Enantioseparation, Banerjee_2020_JACS_enantioseparation, larionov2021enantioselective, Lu_2021_physchemlett_enantioseparation, Bhowmick_2021_crystalgrowth_enantioseparation, Ozturk_2023_Sciadv_enantioseparation, Lu_2023_JPhysChemC_enantioseparation, yu2023advances} 

Recently, the authors and their collaborators have proposed that the electric toroidal monopole, a time-reversal-even pseudoscalar, serves as a microscopic order parameter to quantify chirality in electronic systems~\cite{Hayami_PhysRevB.98.165110, kishine2022definition, kusunose2024emergence}. 
Within the framework of multipole representation theory, which unifies four classes of multipoles (electric, magnetic, electric toroidal, and magnetic toroidal multipoles)~\cite{kusunose2022generalization, hayami2024unified}, chirality has been identified as a degree of freedom emerging from the entanglement of spin, orbital, and lattice degrees of freedom. 
This theoretical advance provides a powerful tool for analyzing the microscopic origins of electron-driven chiral orderings, which have been proposed in systems such as Cd$_2$Re$_2$O$_7$~\cite{yamaura2002low, Matteo_PhysRevB.96.115156, hiroi2018pyrochlore, Hayami_PhysRevLett.122.147602, hirai2022successive}, URu$_2$Si$_2$~\cite{Kambe_PhysRevB.97.235142, kambe2020symmetry, hayami2023chiral}, URhSn~\cite{harima2023hidden, kusunose2024configuration, tabata2025successive, ishitobi2025purely}, and Nd$_3$Rh$_4$Sn$_{13}$~\cite{Shimoda_PhysRevB.109.134425}, in conjunction with analyses based on (magnetic) point group theory~\cite{Hayami_PhysRevB.98.165110, Yatsushiro_PhysRevB.104.054412, Cheong2025chirality}. 
Furthermore, this framework goes beyond the conventional symmetry-based classification of chirality, enabling its quantitative treatment directly from microscopic electronic wave functions. 
Such a perspective has recently been reinforced by first-principles studies of a twisted methane molecule~\cite{inda2024quantification} and elemental Te~\cite{oiwa2025predominant}. 
For the latter case, it was demonstrated that inter-$p$-orbital hoppings along the helical lattice structure play a decisive role in stabilizing chiral lattice structures. 
Together, these developments pave the way for a unified microscopic description of chirality in solids, bridging abstract symmetry considerations with realistic electronic-structure mechanisms.

In the present study, we explore further cross-correlation effects associated with electronic chirality. 
Our focus is on the effective coupling, in chiral systems, between time-reversal-even polar- and axial-vector electronic degrees of freedom. 
The former is described by an electric dipole, while the latter is described by an electric toroidal dipole. 
Here, the electric toroidal dipole acts as a microscopic order parameter of axial systems that preserve both spatial inversion and time-reversal symmetries, as recently observed in materials such as RbFe(MoO$_4$)$_2$~\cite{jin2020observation, Hayashida_PhysRevMaterials.5.124409}, NiTiO$_3$~\cite{hayashida2020visualization, Hayashida_PhysRevMaterials.5.124409, yokota2022three, Bhowal_PhysRevResearch.6.043141}, Ca$_5$Ir$_3$O$_{12}$~\cite{Hasegawa_doi:10.7566/JPSJ.89.054602, hanate2021first, hayami2023cluster, hanate2023space}, and K$_2$Zr(PO$_4$)$_2$~\cite{yamagishi2023ferroaxial, Bhowal_PhysRevResearch.6.043141}. 
Such an electric toroidal dipole has also attracted considerable attention as a driver of unconventional cross-correlation responses between conjugate fields and physical observables~\cite{Hayami_doi:10.7566/JPSJ.91.113702}. 
Examples include the antisymmetric thermopolarization~\cite{Nasu_PhysRevB.105.245125}, intrinsic longitudinal spin current generation~\cite{Roy_PhysRevMaterials.6.045004, Hayami_doi:10.7566/JPSJ.91.113702}, transverse nonlinear magnetic responses~\cite{inda2023nonlinear}, unconventional Hall effect~\cite{Hayami_PhysRevB.108.085124}, and second-order nonlinear magnetostriction~\cite{kirikoshi2023rotational}. 

Building on this coupling, we theoretically demonstrate the emergence of chirality-induced axiality at surface polarization (CIAS), whereby the handedness of chirality of substrates energetically favors an axial moment with a specific orientation at the surface. 
Conversely, we establish the existence of the inverse process, axiality-induced chirality at surface polarization (AICS), in which an axial moment of substrates with a given orientation dictates the preferred handedness of chirality.
These reciprocal couplings between chirality and axiality define a new paradigm of symmetry-governed functionalities, providing a microscopic route to achieve single-domain states of chirality and axiality. 
Moreover, they suggest promising avenues for selective adsorption and crystallization of chiral and axial molecules, with potential implications for spintronics, molecular electronics, and enantioselective catalysis.
Note that Ghosh {\it et al.} first proposed enantioselective interaction between chiral molecules and a ferromagnetic substrate~\cite{Koyel_2018_science_enantioseparation}, leading to enantioselective absorption~\cite{Banerjee_2020_JACS_enantioseparation, Lu_2021_physchemlett_enantioseparation, miwa2024spin} and crystallization~\cite{Tassinari_ChemSci_2019_Enantioseparation, Bhowmick_2021_crystalgrowth_enantioseparation, Ozturk_2023_Sciadv_enantioseparation} of chiral molecules.
In contrast to this magnetic interaction, we propose a distinct non-magnetic mechanism: a coupling between chirality and axiality mediated by the polar surface of a non-magnetic substrate.

\begin{figure}[tb!]
\begin{center}
\includegraphics[width=1.0\hsize]{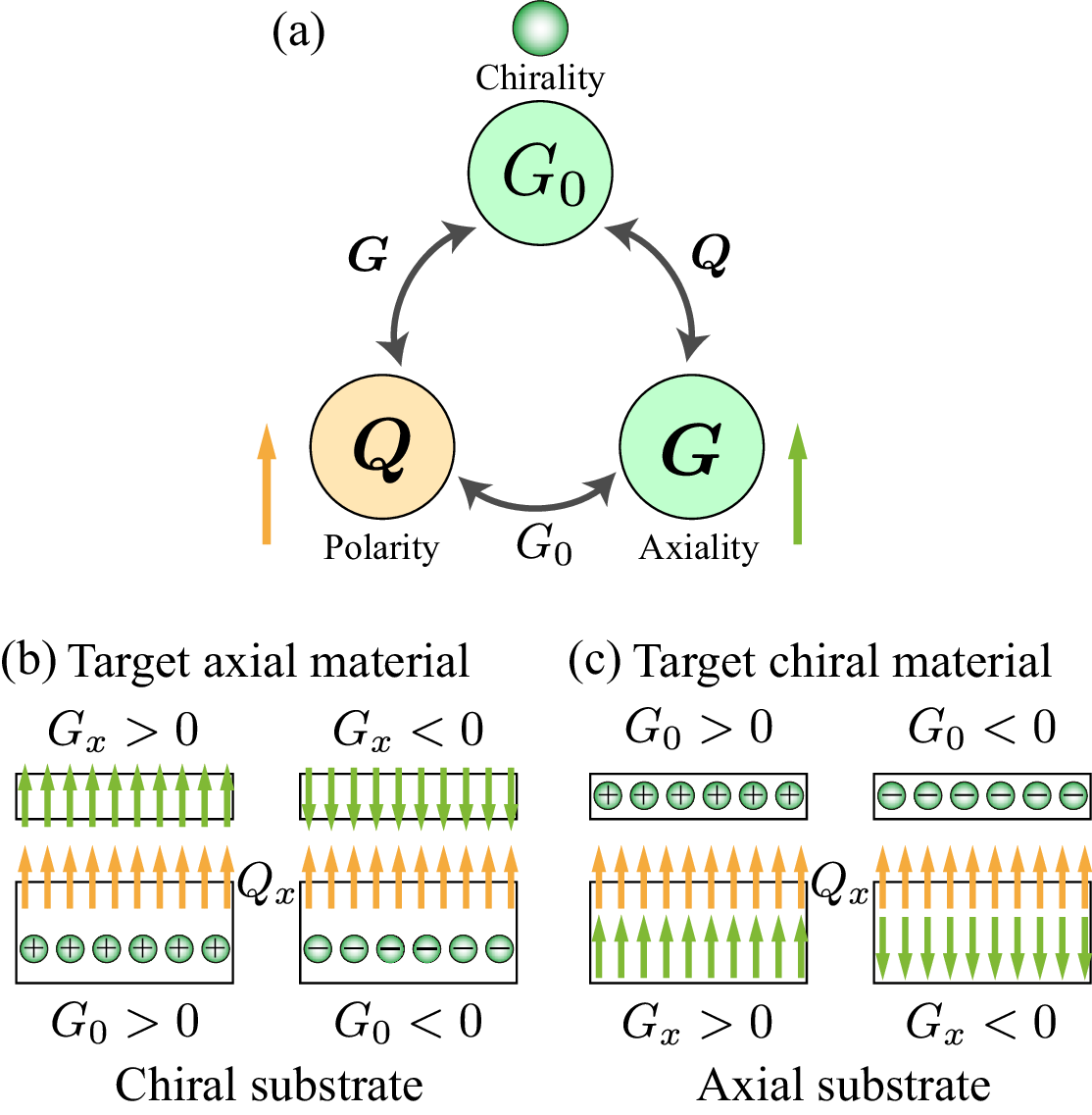} 
\caption{
\label{fig: ponti} 
(Color online) (a) Mutual relationships between the electric toroidal monopole $G_0$, electric toroidal dipole $\bm{G}$, and electric dipole $\bm{Q}$. 
(b,c) Schematic picture of single-domain selection in terms of (b) axial systems on the chiral substrate and (c) the chiral systems on the axial substrate along the $x$ direction. 
The green spheres, green arrows, and orange arrows stand for $G_0$, $G_x$, and $Q_x$. 
}
\end{center}
\end{figure}

We begin by examining the symmetry relationships among chirality, axiality, and polarity.  
From the viewpoint of symmetry and multipole representation, chirality is characterized by the rank-0 electric toroidal monopole $G_0$ (a time-reversal-even pseudoscalar), axiality is described by the rank-1 electric toroidal dipole $\bm{G}=(G_x, G_y, G_z)$ (a time-reversal-even axial vector), and polarity is represented by the rank-1 electric dipole $\bm{Q}=(Q_x, Q_y, Q_z)$ (a time-reversal-even polar vector). 
These three multipoles are mutually connected through the relations
\begin{align}
\label{eq: G0}
G_0 &\leftrightarrow \bm{Q}\cdot \bm{G}, \\
\bm{G} &\leftrightarrow G_0 \bm{Q}, \\
\bm{Q} &\leftrightarrow G_0 \bm{G},  
\end{align}
as schematically illustrated in Fig.~\ref{fig: ponti}(a). 
This correspondence implies that the emergence of any one of the three multipoles induces correlations with the other two, providing a natural framework to interpret cross-correlation phenomena under these ordered states. 
For instance, in a chiral ($G_0$-active) system, the coupling between $\bm{Q}$ and $\bm{G}$ underlies the electric-field-induced rotational distortion~\cite{Oiwa_PhysRevLett.129.116401}. 
Conversely, in an axial ($\bm{G}$-active) system, the interplay between $\bm{Q}$ and $G_0$ manifests as an electric-field-induced optical rotation, i.e., electrogyration effect~\cite{hayashida2020visualization, Martinez2025ferroaxial}. 
In this manner, chirality, axiality, and polarity are fundamentally entangled with one another. 

The interplay among these multipoles naturally raises the question of whether the signs of $G_0$ and $\bm{G}$ can be energetically controlled---namely, whether single-domain states can be realized in chiral and axial systems. 
To address this issue, we theoretically introduce the concepts of the CIAS and AICS effects, analogous to the CISS effect. 
Since the polar field corresponding to $\bm{Q}$ is inherently present at a surface perpendicularly, it couples to $\bm{G}$ ($G_0$) in a chiral (axial) substrate with finite $G_0$ ($\bm{G}$). 
On this basis, the handedness of chirality, i.e., the sign of the expectation value of $G_0$, is directly linked to the orientation of the axial moment, i.e., the sign of the expectation value of $\bm{G}$, via the surface field $\bm{Q}$. 
Specifically, a state with $G_x>0$ ($G_x<0$) is energetically favored when placed on a chiral substrate with $G_0>0$ ($G_0<0$) under $Q_x>0$, corresponding to the CIAS effect, as illustrated in Fig.~\ref{fig: ponti}(b). 
Conversely, when a chiral material is placed on an axial substrate, the sign of $G_0$ is determined by the orientation of the axial moment $G_x$, giving rise to the AICS effect, as shown in Fig.~\ref{fig: ponti}(c). 
In the following, we demonstrate these two reciprocal effects by analyzing a minimal tight-binding model, thereby establishing a microscopic basis for the energetic selection of chirality and axiality at surfaces.  

\begin{figure}[tb!]
\begin{center}
\includegraphics[width=1.0\hsize]{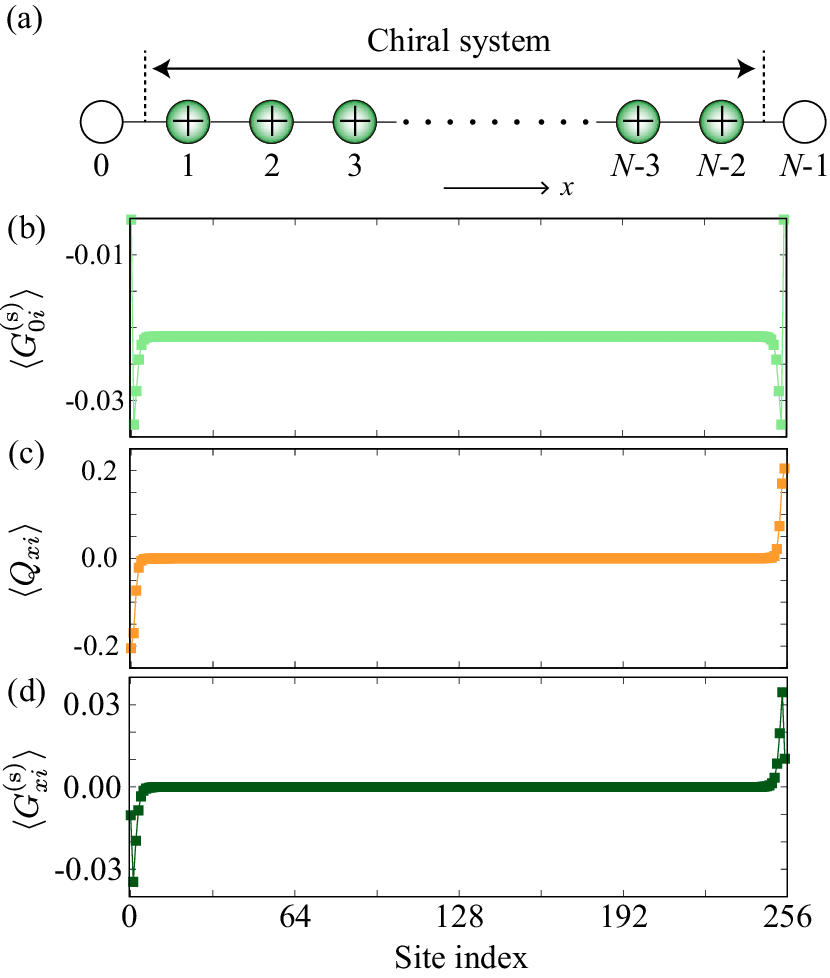} 
\caption{
\label{fig: fig2} 
(Color online) (a) Chiral system interfaced with non-chiral sites. 
(b-d) Expectation values of (b) the electric toroidal monopole $G^{\rm (s)}_0$, (c) electric dipole $Q_x$, and (d) electric toroidal dipole $G^{\rm (s)}_x$ per site $i$. 
}
\end{center}
\end{figure}

To extract a minimal essence, we adopt the $s$-$p$ hybridized tight-binding model, which includes the electric toroidal monopole (chirality), electric toroidal dipole (axiality), and electric dipole (polarity) degrees of freedom in the Hilbert space, on a one-dimensional chain along the $x$ direction, as illustrated in Fig.~\ref{fig: fig2}(a). 
The Hamiltonian is given by 
\begin{align}
\label{eq: Ham}
\mathcal{H}&= \sum_{ij\alpha\alpha' \sigma}(t^{\alpha \alpha'}_{ij} c^{\dagger}_{i\alpha\sigma} c^{}_{j\alpha'\sigma} + {\rm H.c.})+\frac{\lambda}{2} \sum_{i \tilde{\alpha} \tilde{\alpha}' \sigma \sigma'} c^{\dagger}_{i \tilde{\alpha} \sigma} (\bm{l}\cdot \bm{\sigma})_{\tilde{\alpha}\tilde{\alpha}'}^{\sigma\sigma'} c^{}_{i \tilde{\alpha}' \sigma'} \nonumber \\
&+  \sum_{i\alpha\alpha' \sigma\sigma'}h^i_{G_0} c^{\dagger}_{i\alpha\sigma}[G^{\rm (s)}_0]_{\alpha\alpha'}^{\sigma\sigma'} c^{}_{i\alpha'\sigma'}
+  \sum_{i\alpha\alpha' \sigma\sigma'}h^i_{G_x} c^{\dagger}_{i\alpha\sigma}[G^{\rm (s)}_{x}
]_{\alpha\alpha'}^{\sigma\sigma'} c^{}_{i\alpha'\sigma'}. 
\end{align}
Here, we use $c^{\dagger}_{i\alpha\sigma}$ and $c_{i\alpha\sigma}$ to denote creation and annihilation operators for an electron with spin $\sigma$ in orbital $\alpha=s$, $p_x$, $p_y$, and $p_z$ (or $\tilde{\alpha}=p_x$, $p_y$, and $p_z$) at site $i$. 
The first term includes onsite potential $\Delta$ for the $s$ orbital and nearest-neighbor hoppings: $t_s$ between $s$ orbitals, $t_{x}$ between $p_x$ orbitals, $t_z$ between $p_z$ orbitals, and $t_{sp}$ between $s$ and $p_x$ orbitals. 
The second term represents the atomic spin--orbit coupling for the $p$ orbitals, where $\bm{l}$ and $\bm{\sigma}$ stand for the orbital angular momentum and Pauli matrix in spin space, respectively. 

The third and fourth terms denote the mean field originating from the chiral or axial symmetry. 
The expressions of $G^{\rm (s)}_0$ and $G^{\rm (s)}_x$ are given by~\cite{kusunose2020complete} 
\begin{align}
G^{\rm (s)}_0&= \frac{1}{\sqrt{3}} (T_x\sigma_x + T_y \sigma_y + T_z \sigma_z), \\
G^{\rm (s)}_x&= \frac{1}{\sqrt{2}} (l_y \sigma_z-l_z \sigma_y),
\end{align}
where $\bm{T}=(T_x, T_y, T_z)$ represents the atomic-scale magnetic toroidal dipole arising from the imaginary $s$-$p$ hybridization. 
Thus, $G^{\rm (s)}_0$ corresponds to the inner product between time-reversal-odd polar ($\bm{T}$) and axial ($\bm{\sigma}$) vectors, whereas $G^{\rm (s)}_x$ corresponds to the outer product between orbital and spin angular momenta. 
The quantities $h^i_{G_0}$ and $h^i_{G_x}$ are the magnitudes of the mean field for $G^{\rm (s)}_0$ and $G^{\rm (s)}_x$, respectively. 
Microscopically, $h^i_{G_0}$ can originate from a chiral lattice structure~\cite{oiwa2025predominant} or spontaneous orbital hybridization/order~\cite{hayami2023chiral, ishitobi2025purely}, while $h^i_{G_x}$ can arise from the combined effect of vertical-mirror-symmetry-breaking orbital hybridization and spin--orbit coupling~\cite{Inda_PhysRevB.111.L041104} or from vortex-type spin configurations~\cite{Hayami_PhysRevB.106.144402}. 
Although we evaluated these spin-dependent electric toroidal multipoles, $G_0^{\mathrm{(s)}}$ and $G_x^{\mathrm{(s)}}$, based on a minimal tight-binding model with the atomic spin--orbit coupling, the following results can be applied to the spin-independent electric toroidal multipoles in non-relativistic situations~\cite{ishitobi2025purely, oiwa2025predominant}. 

In the following, we set the model parameters as $t_s=-1$, $t_x=0.8$, $t_z=0.4$, $t_{sp}=0.5$, $\Delta=-3$, and $\lambda=0.6$.
We study a chain of length $N=256$ under open boundary conditions, with sites labeled $i=0$ to $N-1$. 
For clarity, the system is partitioned into the bulk, defined by $1\leq i \leq N-2$, and the two surfaces at $i=0$ and $N-1$. 

First, we examine the case in which the bulk is chiral but the surfaces are non-chiral. 
This is realized by setting $h^i_{G_0}=h_{G_0}=0.5$ for $1\leq i \leq  N-2$, $h^i_{G_0}=0$ for $i = 0, N-1$, and $h^i_{G_x}=0$ for all sites, as illustrated in Fig.~\ref{fig: fig2}(a). 
Figure~\ref{fig: fig2}(b) shows the site-resolved expectation value of $G^{\rm (s)}_0$, $\langle G^{\rm (s)}_{0i} \rangle$, at the chemical potential $\mu=0.3$. 
While $\langle G^{\rm (s)}_{0i} \rangle$ is almost uniform in the bulk, it exhibits oscillations near the surfaces. 

In the surface region, the absence of mirror symmetry with respect to the $yz$ plane induces the electric dipole $Q_{x}$, as shown in Fig.~\ref{fig: fig2}(c). 
Here, the electric dipole $Q_x$ corresponds to the real on-site $s$-$p_x$ hybridization. 
Because the polar field direction is reversed at the two surfaces ($i=0$ and $i=N-1$), the induced $\langle Q_{xi} \rangle $ has equal magnitude but opposite sign at the two ends. 
Through the coupling in Eq.~(\ref{eq: G0}), this nonzero $\langle Q_{xi} \rangle $ generates the axial moment $\langle G^{\rm (s)}_{xi} \rangle$. 
Indeed, Fig.~\ref{fig: fig2}(d) shows that $\langle G^{\rm (s)}_{xi} \rangle$ is induced with opposite signs around the two surfaces.

Such surface-induced $\langle G^{\rm (s)}_x \rangle$ is crucial in determining the energetically preferred orientation of the axial moment when interfaced with an axial system, as will be discussed below. 
We note that $\sum_i \langle Q_{xi} \rangle=\sum_i \langle G^{\rm (s)}_{xi} \rangle = 0$, since $G^{\rm (s)}_x$ does not belong to the identity irreducible representation in the total system. 
Finally, when the sign of $h^i_{G_0}$ is reversed, $\langle G^{\rm (s)}_{0i} \rangle$ changes sign, and then $\langle G^{\rm (s)}_{xi} \rangle$ also changes sign, whereas $\langle Q_{xi} \rangle $ remains unchanged.

\begin{figure}[tb!]
\begin{center}
\includegraphics[width=1.0\hsize]{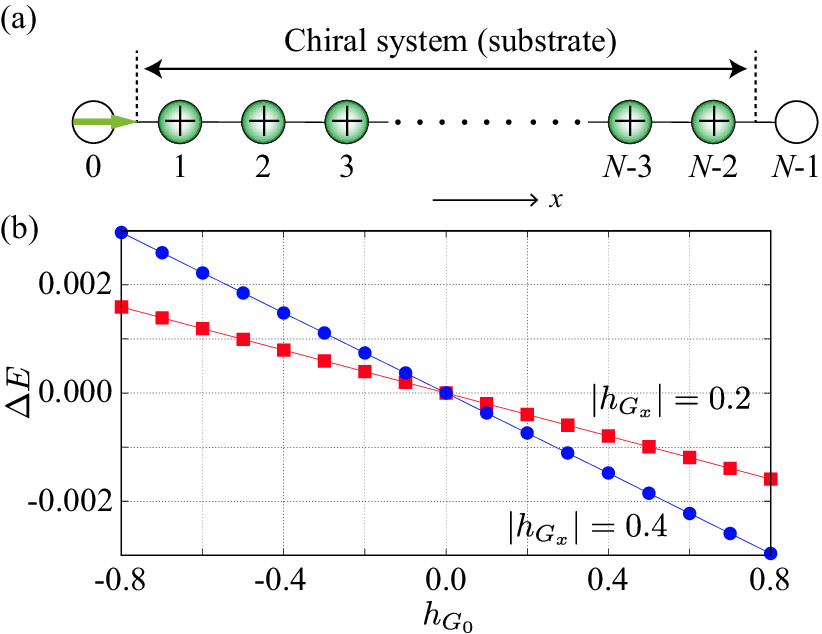} 
\caption{
\label{fig: fig3} 
(Color online) 
(a) Chiral substrate interfaced with an axial site at $i=0$. 
(b) $h_{G_0}$ dependence of the energy difference between the states with $h_{G_x}<0$ and $h_{G_x}>0$, $\Delta E = E(h_{G_x}>0) - E(h_{G_x}<0)$, for different values of $h_{G_x}$ at $\mu=0.3$. 
}
\end{center}
\end{figure}

Next, we turn to the interface system by additionally considering an axial site at $i=0$, as schematically illustrated in Fig.~\ref{fig: fig3}(a). 
We set $h^i_{G_x}=h_{G_x}=\pm 0.2$ or $\pm 0.4$ at $i=0$ ($h^i_{G_x} = 0$ at $i\neq 0$) and examine the $h_{G_0}$ dependence of the total energy difference between states with $h_{G_x}<0$ and $h_{G_x}>0$ at $\mu=0.3$, as shown in Fig.~\ref{fig: fig3}(b). 
The results clearly demonstrate that the energy difference arises depending on the sign of the mean field at the axial site: the configuration with $h_{G_x}>0$ and $\langle G^{\rm (s)}_x \rangle<0 $ is energetically favored compared to that with $h_{G_x}<0$ and $\langle G^{\rm (s)}_x\rangle >0 $ when $h_{G_0}>0$. 
Together with the result at $h_{G_0}=0.5$ in Fig.~\ref{fig: fig2}(d), which shows that $\langle G^{\rm (s)}_{x0} \rangle <0$ is realized at the edge ($i=0$), this indicates the emergence of coupling between the bulk chirality and the surface axial moment. 
Consequently, when the system is placed on a chiral substrate, the axial state with $\langle G^{\rm (s)}_{x} \rangle < 0$ is energetically stabilized, which is precisely the CIAS effect. 
It is noteworthy that the opposite end exhibits reversed axiality selection, reflecting the contrasting trends at the two ends, as shown in Fig.~\ref{fig: fig2}(d). 

When the sign of $h_{G_0}$ is reversed, the stabilization tendency is also reversed, meaning that the handedness of chirality dictates the preferred axial domain state.
Furthermore, the magnitude of the energy splitting grows with increasing $h_{G_0}$ and $h_{G_x}$ [Fig.~\ref{fig: fig3}(b)], implying that materials with intrinsically large $\langle G^{\rm (s)}_{0} \rangle$ and $\langle G^{\rm (s)}_{x} \rangle$ are favorable for realizing robust CIAS. 
Since quantities like $\langle G^{\rm (s)}_{0} \rangle$ and $\langle G^{\rm (s)}_{x} \rangle$ in real materials can be evaluated within the symmetry-adapted closest Wannier formalism~\cite{oiwa2025symmetry}, their systematic assessment may provide a practical guideline for identifying and designing materials that exhibit the giant CIAS effect. 
In this context, it might be useful to combine the calculations based on different quantifications of chirality~\cite{Miki_PhysRevLett.134.226401}.
Finally, the favored sign of the axial moment is not universal but depends on the underlying electronic state and model parameters. 
For example, at $\mu=-0.8$, where the positive axial moment is induced at the edge, the axial moment with $\langle G^{\rm (s)}_{x} \rangle>0$ is energetically favored for $h_{G_0}>0$ (not shown). 

\begin{figure}[tb!]
\begin{center}
\includegraphics[width=1.0\hsize]{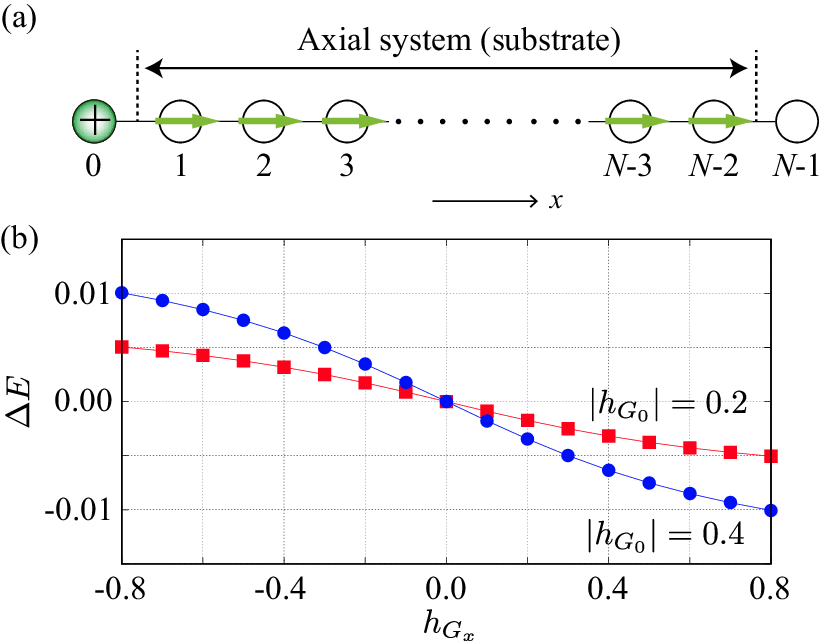} 
\caption{
\label{fig: fig4} 
(Color online) 
(a) Axial substrate interfaced with a chiral site at $i=0$. 
(b) $h_{G_x}$ dependence of the energy difference between the states with $h_{G_0}>0$ and $h_{G_0}<0$, $\Delta E = E(h_{G_0}>0) - E(h_{G_0}<0)$, for different values of $h_{G_0}$ at $\mu=-1.4$. 
}
\end{center}
\end{figure}

Analogous to the CIAS effect, we demonstrate its inverse counterpart, namely, the AICS effect. 
To this end, we interchange the roles of the chiral and axial mean fields in Fig~\ref{fig: fig3}(a); we set $h^i_{G_x}=h_{G_x}\neq 0$ for $1\leq i \leq  N-2$, $h^i_{G_x}=0$ for $i = 0, N-1$, $h^i_{G_0}=h_{G_0}\neq 0$ for $i= 0$, and $h^i_{G_0}=0$ for $1\leq i \leq N-1$. 
In other words, the bulk acts as an axial medium, whereas the surfaces play the role of chiral sites. 
Equivalently, the situation can be interpreted as a chiral site residing on an axial substrate.

In this configuration, $\langle G^{\rm (s)}_{xi} \rangle$ is induced across the system, while $\langle G^{\rm (s)}_{0i} \rangle$ appears only near the surfaces, with opposite signs at the two ends. 
Figure~\ref{fig: fig4}(b) displays the $h_{G_x}$ dependence of the total energy difference between the states with $h_{G_0}>0$ and $h_{G_0}<0$ at $\mu=-1.4$. 
As in the results of Fig.~\ref{fig: fig3}(b), the energy difference is governed by the sign reversal of $h_{G_0}$, corresponding precisely to the AICS effect. 
As in the case of CIAS, the energetically preferred chiral domain depends sensitively on the underlying electronic states and model parameters. 
For instance, at $\mu=-1.8$, the favored sign of $\langle G^{\rm (s)}_0 \rangle$ is reversed compared with the $\mu=-
1.4$ case (not shown). 
Furthermore, reversing the sign of $h_{G_x}$ flips the stabilization tendency, such that the orientation of the axial moment selects an alternative chiral domain.

Our results demonstrate that CIAS and its inverse, AICS, provide powerful routes for engineering single-domain states and interfacial functionalities in solids. 
Because a polar field $\bm{Q}$ is inherently present at any surface perpendicularly, axiality $\bm{G}$ (chirality $G_0$) naturally emerges at the surface of chiral (axial) systems with finite $G_0$ ($\bm{G}$), through the coupling shown in Fig.~\ref{fig: ponti}. 
This establishes CIAS and AICS as surface-driven mechanisms for selecting the sign of axiality or chirality, thereby enabling the stabilization of single-domain states without the need for bulk symmetry breaking or external fields.

The energy splitting between competing domain configurations grows with the interfacial chiral and axial mean fields ($h_{G_0}$ and $h_{G_x}$), indicating that materials with large $\langle G^{\rm (s)}_0 \rangle$ and $\langle G^{\rm (s)}_x \rangle$, as well as interface designs with enhanced surface polarity $Q_x$, are promising candidates for amplifying CIAS and AICS. 
In addition, electrostatic gating, surface termination, or polar-layer insertion can be exploited to tune $\bm{Q}$, providing versatile means of controlling these effects.

The selection rules associated with CIAS and AICS also highlight potential applications in molecular systems. 
At molecule-solid interfaces, they manifest as enantioselective adsorption on axial surfaces and axial-selective adsorption on chiral surfaces, offering new pathways for chiral separation, interfacial catalysis, and molecular recognition governed by the sign of the interfacial multipole. 
Such functionalities open promising opportunities for practical applications in spintronics, molecular electronics, and enantioselective technologies.

To summarize, we have introduced two reciprocal interfacial selection mechanisms---\textit{chirality-induced axiality at surface polarization} (CIAS) and \textit{axiality-induced chirality at surface polarization} (AICS)---that convert the ubiquitous surface polarity into deterministic control over the signs of $G_0$ (chirality) and $\bm{G}$ (axiality). 
Using a minimal $s$-$p$ hybridized chain with spin--orbit coupling, we showed that the surface polar field $\bm{Q}$ couples intrinsically to $G_0$ and $\bm{G}$, giving rise to (i) CIAS at a chiral substrate/axial site interface, where the induced $\langle G^{\rm (s)}_x\rangle$ near the surfaces biases the axial domain and yields an energy splitting controlled by $h_{G_0}$, and (ii) the inverse AICS at an axial substrate/chiral site interface, where $h_{G_x}$ (and $G_x$) determines the sign of $G_0$. 
These biases are reversible by flipping the interfacial fields and tunable through electronic filling, enabling voltage-, termination-, or layer-controlled domain selection without bulk symmetry breaking or external magnetic fields.

The magnitude of the energy splitting is related to the magnitude of $h_{G_0}$ and $h_{G_x}$, and the surface polarity $|\bm{Q}|$, providing practical design knobs: materials with large $\langle G^{\rm (s)}_0\rangle$ and $\langle G^{\rm (s)}_x\rangle$, polar terminations, and gate-tunable interfaces. 
Beyond domain control, the same selection rules apply directly to molecule-solid interfaces, enabling enantioselective adsorption on axial surfaces (AICS) and axial-selective adsorption on chiral surfaces (CIAS). 
These reciprocal effects broaden the scope of symmetry-governed functionalities and pave the way for advanced applications in chiral separation, interfacial catalysis, molecular recognition, spintronics, and molecular electronics.

\begin{acknowledgments}
This research was supported by JSPS KAKENHI Grants Numbers JP22H00101, JP22H01183, JP23H04869, JP23K03288, and by JST CREST (JPMJCR23O4) and JST FOREST (JPMJFR2366). 
Parts of the numerical calculations were performed in the supercomputing systems in ISSP, the University of Tokyo.
\end{acknowledgments}

\appendix

\bibliographystyle{jpsj}
\bibliography{ref.bib}

\end{document}